# An efficient coding algorithm for general Framed Pulse Width Modulations

Soon-Won Kwon, Hyeon-Min Bae

*Abstract*— This paper introduces a new coding algorithm for Framed Pulse Width Modulation (FPWM). The proposed algorithm requires 93% fewer look-up tables (LUTs) than the previous FPWM coding algorithm and increases a bitrate by 25%. The proposed algorithm is compatible with general FPWM with various frame lengths and pulse width resolutions. Theoretical bitrates and the sizes of LUT required for coding various FPWMs are also provided. The MATLAB simulation demonstrates the proposed FPWM signal which contains 14-bit information in 8 UI frame length, showing 75% higher bitrate than the NRZ signal with the same baud rate. The decoding algorithm restores the original bit without any bit error and validates the proposed FPWM and its coding scheme.

## I. INTRODUCTION

Despite the ever-increasing data rate demands, the channel capacity of copper cables suffers fundamental bandwidth limitation. Such limited bandwidth is leading the industry to adopt PAM4 modulation for data rates above 56Gbps.[1-8] However, tight linearity requirements for multi-level signaling techniques make it difficult to meet low BER conditions and consequently impede rapid application of PAM4 in various industries.[3,4,6-8] To alleviate this linearity issue, a Framed Pulse Width Modulation (FPWM) technique was introduced[9]. The FPWM transmits bit data by modulating the edge phase of a binary signal. However, the previous FPWM coding scheme, which relies solely on a look-up table (LUT), is not suitable for more complex FPWMs with higher bitrates, since the size and complexity of LUT increase exponentially as the frame length increases. Therefore, this paper introduces a new FPWM scheme and its coding algorithm that reduces the size of the LUT by 93% and increases the bitrate by 25% compared to the previous FPWM. The proposed algorithm can be applied to general FPWMs with various frame lengths and pulse width resolutions.

The remainder of this paper is organized as follows: Section II.A explains the concept of general FPWM and its theoretical bitrate. Section II.B discusses the encoding and decoding algorithms for the proposed FPWM scheme. In Section III, MATLAB simulation results are presented to verify the feasibility of proposed algorithms. Section IV is the conclusion.

Soon-Won Kwon, SeJun Jeon, Woohyun Kwon, Bongjin Kim, Gain Kim and Hyeon-Min Bae are with the school of Electrical Engineering, Korea Advanced Institute of Science and Technology(KAIST), Daejeon, Korea (e-mail: ksw8538@kaist.ac.kr; jsjwin@kaist.ac.kr; dngusrnjs@kaist.ac.kr; bjkim3927@kaist.ac.kr; gikim@kaist.ac.kr; hmbae@kaist.ac.kr).

## II. GENERAL FRAMED PULSE-WIDTH MODULATIONS (FPWM)

### A. Theoretical bitrate calculation

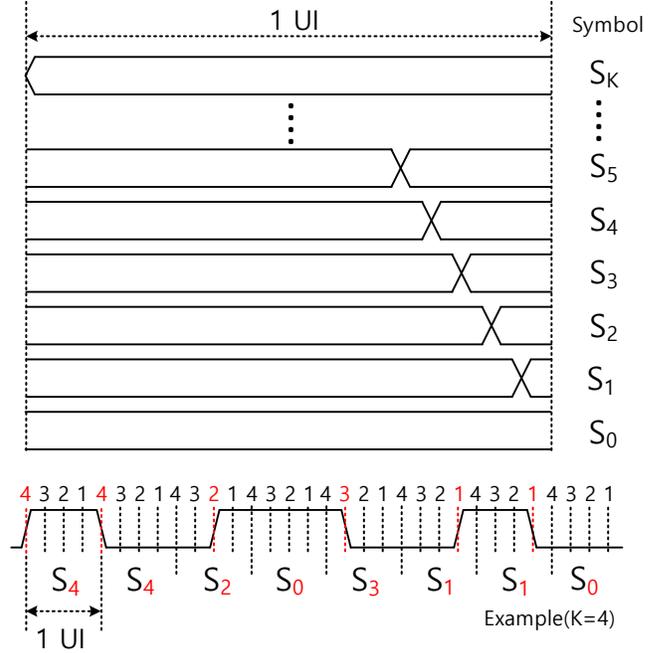

Fig. 1. *K+1* FPWM symbols with pulse width resolution *K*

FPWM encodes binary data into the edge phases of transmitted signals. As shown in Fig.1, symbols with edges at $K$ different phases within 1 UI period are denoted as '$S_1$', '$S_2$', '$S_3$', ... '$S_K$', respectively. If there is no signal transition, the symbol is denoted as '$S_0$'. The rising or falling state of the symbol depends only on the state of the previous symbol and is independent of the encoded data. The symbols in Fig.1 are combined into one frame of $m$ UI length. If the number of possible combinations of symbol arrays in a frame is $N$, then the maximum number of binary data that can be carried by one frame is $\lfloor \log_2 N \rfloor$. Since only an integer number of bits can be encoded, the floor function ($\lfloor \ \rfloor$) is applied.

The pulse width of the FPWM signal must be at least 1 UI to maintain the same bandwidth as NRZ. This constraint on the minimum pulse width requires some rules between adjacent symbols as shown in Fig.2. If a single pulse created by combining two consecutive symbols has a width equal to or larger than 1 UI, the next symbol after '$S_q$' must be '$S_0$'~ '$S_q$'. However, all symbol can be placed after '$S_0$' since the pulse width is already greater than 1 UI. Therefore, only '$S_K$' and '$S_0$' symbols allow all kinds of symbols to be connected behind. On the other hand, FPWM signals are encoded on a frame-by-frame basis, meaning that complete independence between frames must be guaranteed. This independence can be achieved by setting the last symbol of the frame to '$S_K$' or '$S_0$' only.

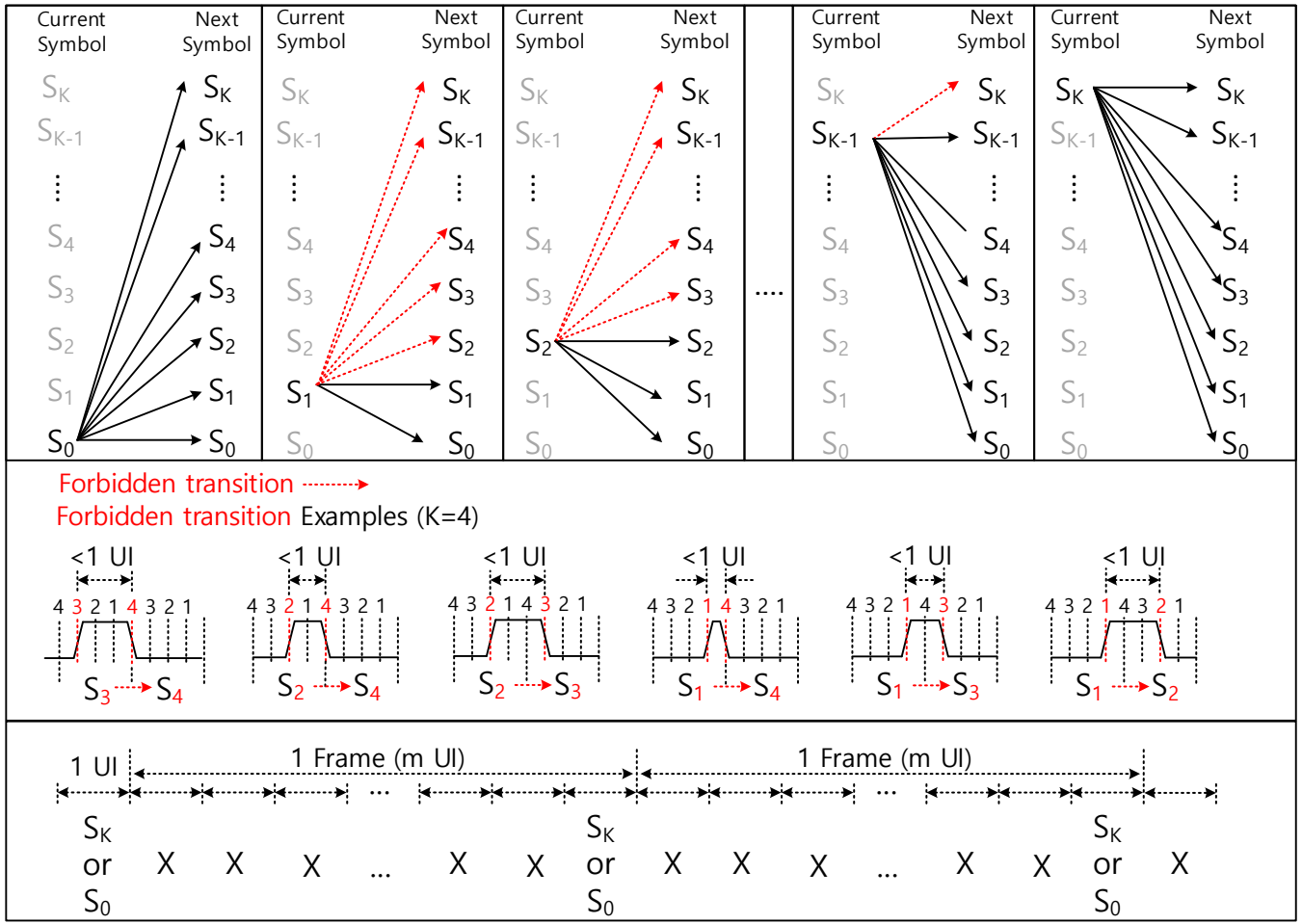

Fig. 2. The coding rules for proposed FPWM

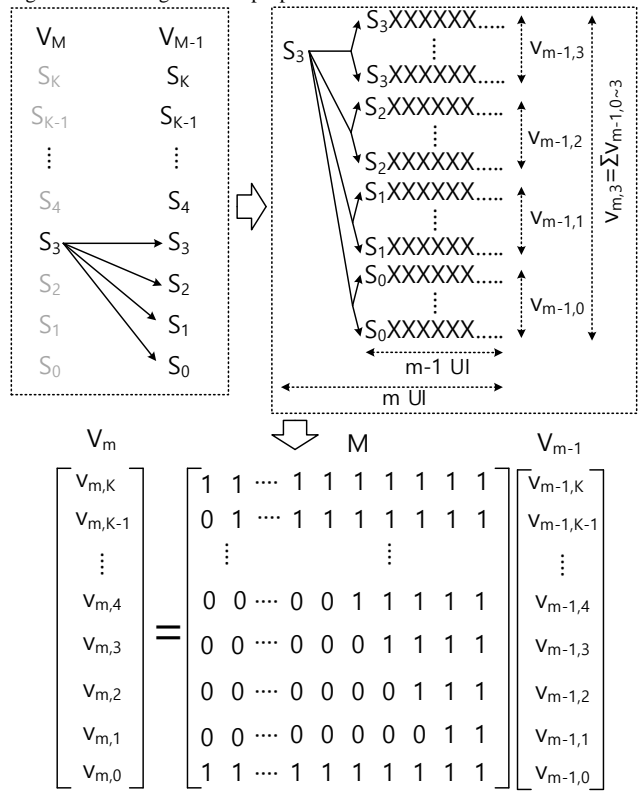

Fig. 3. The bitrate calculation for proposed FPWM

The number of possible symbol arrays in a frame satisfying these rules is derived as shown in Fig.3. The component value $v_{m,q}$ of vector $V_m$ denotes the number of possible combinations of $m$ symbol arrays starting with symbol '$S_q$'. For example, if the frame length is 1 UI, only '$S_0$' or '$S_K$' symbol is possible. Therefore,

$$V_1 = \begin{bmatrix} 1 & 0 & 0 & \ldots & 0 & 1 \end{bmatrix}^T. \quad (1)$$

According to the rules in Fig.2, the number of symbol arrays of length $m$ starting with '$S_q$' is the sum of the number of symbol arrays of length $m-1$ starting with '$S_0$' ~ '$S_q$'. Then, the relationship between $v_{m,q}$ and $v_{m-1,q}$ can be expressed as follows.

$$v_{m,q} = \begin{cases} \sum_{h=0}^{q} v_{m-1,h} & (if \ q > 0) \\ \sum_{h=0}^{K} v_{m-1,h} & (if \ q = 0) \end{cases} \quad (2)$$

Eq.(2) can be expressed in a matrix form, and it is possible to derive $V_m$ from $V_1$ as follows.

$$V_m = \begin{bmatrix} v_{m,K} \\ v_{m,K-1} \\ \vdots \\ v_{m,2} \\ v_{m,1} \\ v_{m,0} \end{bmatrix} = \begin{bmatrix} 1 & 1 & \ldots & 1 & 1 & 1 \\ 0 & 1 & \ldots & 1 & 1 & 1 \\ \vdots & \vdots & \ddots & \vdots & \vdots & \vdots \\ 0 & 0 & \ldots & 1 & 1 & 1 \\ 0 & 0 & \ldots & 0 & 1 & 1 \\ 1 & 1 & \ldots & 1 & 1 & 1 \end{bmatrix} \begin{bmatrix} v_{m-1,K} \\ v_{m-1,K-1} \\ \vdots \\ v_{m-1,2} \\ v_{m-1,1} \\ v_{m-1,0} \end{bmatrix} = MV_{m-1} = M^{m-1}V_1 \quad (3)$$

The total number of symbol arrays that can be represented as one frame of FPWM with $m$ UI length is $N = \sum_{h=0}^{K} v_{m,h}$ so that $\lceil \log_2 N \rceil$ bits can be encoded into the frame. The normalized bitrate of the proposed FPWM scheme with the frame length of *m* UI and pulse width resolution of *K* is calculated as follows.

$$bitrate\,(bit/UI) = \frac{\lfloor \log_2 \sum_{h=0}^{K} v_{m,h} \rfloor}{m} \quad (4)$$

The normalized bitrate means the number of bits encoded per 1 UI. The normalized bitrates of the NRZ, PAM4, and the previously published FPWM scheme are 1, 2, and 1.33 bit/UI, respectively.[9] Fig.4 plots the normalized bitrates of the proposed FPWMs at various frame lengths and pulse width resolutions. It shows that the bitrate of the proposed FPWM at $m = 6$ and $K = 4$ is about 25% higher than the previous FPWM with the same frame length and pulse width resolution.

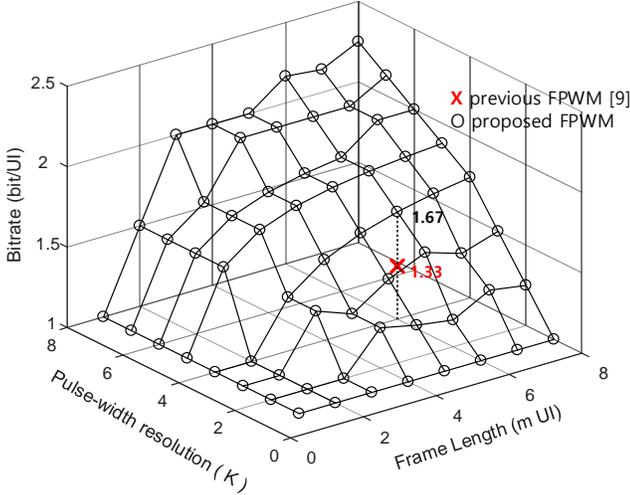

Fig. 4. The calculated bitrates of the proposed FPWMs with various Frame lengths and pulse width resolutions

### B. Coding algorithm

The encoder converts the input binary data into its corresponding FPWM symbol array. The array composed of $K+1$ different symbols ('$S_0$' ~ '$S_K$') can be regarded as a numeral system of radix $K+1$ with Least Significant Digit (LSD) on the right and Most Significant Digit (MSD) on the left. Table I shows an example of FPWM code with $K = 4$ corresponding to the input binary data. If the coding rule in Fig.2 is not applied, the encoder structure can be represented simply as a binary to the base-5 converter as shown in Fig.5. Conversion starts sequentially from the MSD. At each step, the input value is quantized to five levels. The result of the quantization in blue is then removed from the input, and the residual value passes to the next step for further quantization. Each step performs quantization independently of each other, enabling pipeline operation.

TABLE I
CODE TABLE OF THE PROPOSED FPWM ($K$=4)

| Decimal | binary | FPWM (w/o coding rule) | FPWM (with coding rule) |
|---|---|---|---|
| | MSB→LSB | MSD→LSD | MSD→LSD |
| 0 | 00…0000 | $S_0…S_0S_0S_0S_0$ | $S_0…S_0S_0S_0S_0$ |
| 1 | 00…0001 | $S_0…S_0S_0S_0S_1$ | $S_0…S_0S_0S_0S_4$ |
| 2 | 00…0010 | $S_0…S_0S_0S_0S_2$ | $S_0…S_0S_0S_1S_0$ |
| 3 | 00…0011 | $S_0…S_0S_0S_0S_3$ | $S_0…S_0S_0S_2S_0$ |
| 4 | 00…0100 | $S_0…S_0S_0S_0S_4$ | $S_0…S_0S_0S_3S_0$ |
| 5 | 00…0101 | $S_0…S_0S_0S_1S_0$ | $S_0…S_0S_0S_4S_0$ |
| 6 | 00…0110 | $S_0…S_0S_0S_1S_1$ | $S_0…S_0S_0S_4S_4$ |
| 7 | 00…0111 | $S_0…S_0S_0S_1S_2$ | $S_0…S_0S_1S_0S_0$ |
| ⋮ | ⋮ | ⋮ | ⋮ |

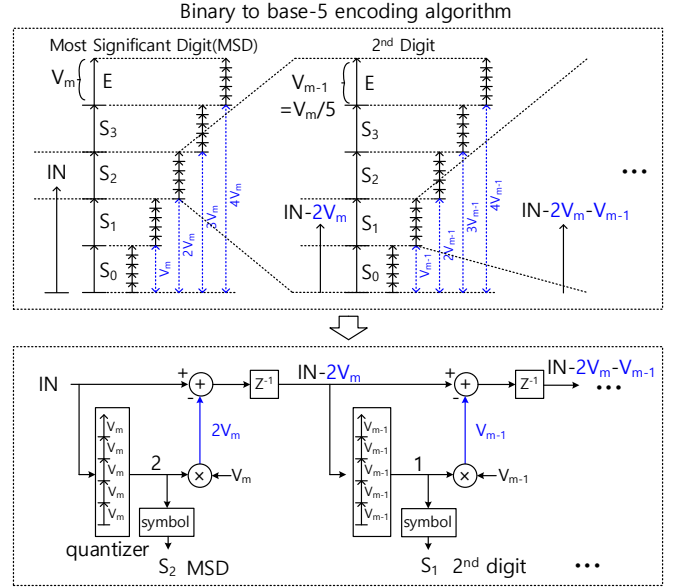

Fig. 5. Block diagram of binary to base-5 converting algorithm

However, if the coding rule in Fig.2 is applied, the forbidden transition should be excluded. Fig.6 shows the modified encoder structure. According to the coding rule, the number of possible symbols in the next digit depends on the quantization result of the current digit. However, by subtracting the quantization result from the input value by the quantity indicated by blue in Fig.6, it is possible to perform independent pipeline encoding operation for each digit. This independent pipeline operation dramatically reduces the size of the LUT for encoding and enables unified FPWM coding theory for various frame lengths and pulse width resolutions.

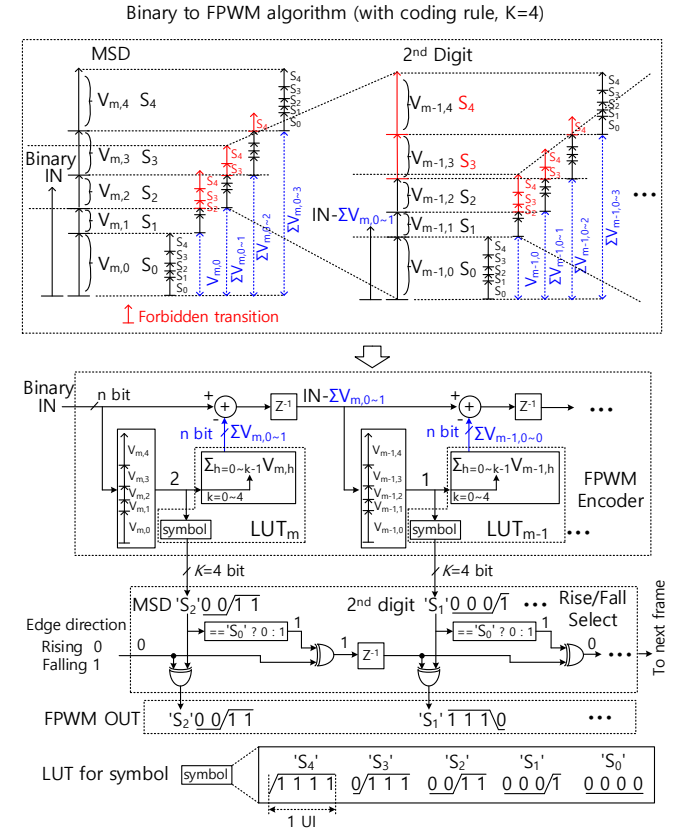

Fig. 6. Block diagram of encoding algorithm for FPWM with $K$=4

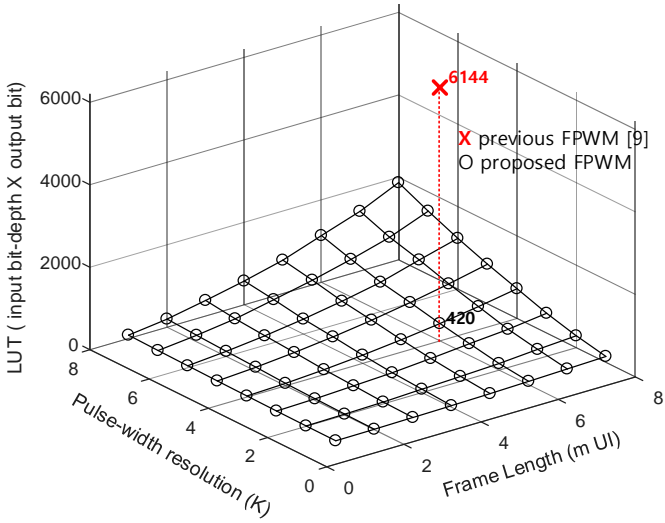

Fig. 7. The calculated LUT size required for encoding the proposed FPWMs with various Frame lengths and pulse width resolutions

The quantized results at each stage of the encoder are converted into FPWM symbols stored in the LUT. Each symbol requires $K$ bits to represent $K$ different edge phases. The symbols stored in the LUT have only rising edges. However, the rising and falling edges must alternate. Therefore, in order to transfer continuous FPWM symbols, the current symbol must be inverted to have the edge direction opposite to that of the previous symbol. The information of the edge direction of the current symbol is transmitted to the subsequent symbol while continuously toggling. This information transfer is performed between frames as well as symbols within a frame. However, if the current symbol is '$S_0$', it means that there is no edge, so the information of the edge direction is transmitted to the next symbol without toggling. Then, the TX driver sends the connected FPWM symbols to the RX side.

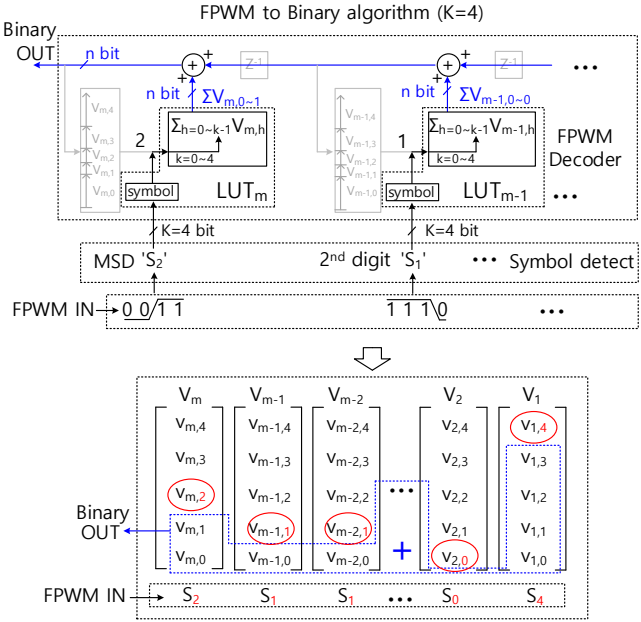

Fig. 8. Block diagram of decoding algorithm for FPWM with K=4

Fig.7 shows the size of LUTs required for encoding the proposed FPWMs of various frame lengths and pulse width resolutions. In the previous coding method, the LUT with $2^8$ input bit-depth outputs 24 bits to convert 8-bit data to one frame of FPWM.[9] On the other hand, the proposed FPWM with the same frame length ($m$=6 UI) and pulse width resolution ($K$=4) can encode 10 bits in one frame. Therefore, as shown in Fig.6, the input bit-depth of the LUT for one digit is 5(=$K+1$), and the LUT output is 14bit (=$n+K$). A total of six LUTs are required, from MSD to LSD. If the size of LUT is estimated as a product of input bit-depth and the number of output bits, the LUT size for the previous FPWM encoding scheme is 6144(=$2^8$×24). However, the size of the LUT required for the proposed coding algorithm is 420(=5×14×6), which is about 93% smaller than the conventional FPWM. Since the LUT is the biggest bottleneck of the previous coding method, the significantly reduced LUT in size enables more complex FPWM coding with higher bitrates.

The decoding algorithm converts the input FPWM symbol array into the original binary data, and it can be implemented by inverting the encoding algorithm as shown in Fig.8. Therefore, the decoder only requires much smaller LUT than conventional methods. This decoding procedure can be understood as an operation of adding all the vector components $v_{m,q}$ below the row corresponding to the received symbol in each $V_m$ vector, as shown in the lower part of Fig.8.

## III. SIMULATION

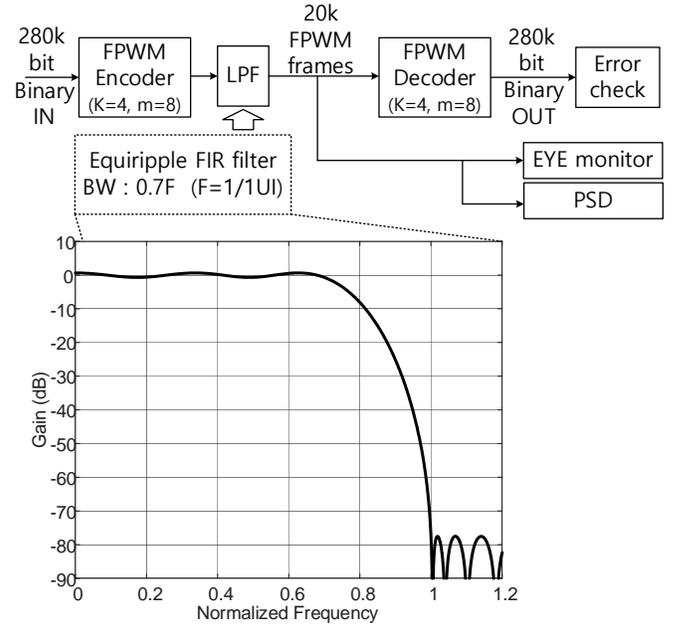

Fig. 9. MATLAB simulation setup

Fig.9 shows the MATLAB simulation setup to verify the proposed FPWM scheme and its coding algorithm. The frame length and the pulse width resolution are set to $m$ = 8 UI and $K$ = 4, respectively. The proposed coding algorithm can encode 14-bit data into 8 UI, so the bitrate is 1.75. A total of 280k bits is converted into 20k frames and then transmitted to the decoder. For comparison, 160k bits of data are also sent as NRZ signals with the same baud rate. Both of the encoded FPWM signal and the NRZ signal pass through an equiripple low pass filter (LPF) with a bandwidth of 0.7$F$, where $F$ is a normalized frequency with a period of one UI. Then, the decoder is fed the LPF output, recovering original binary data without bit errors. The transmitted FPWM signal provides the bitrate 75% greater than NRZ with the same bandwidth.

Fig.10 shows the eye-diagram and power spectral density (PSD) of the encoded FPWM signal. Since the last symbol of the frame is '$S_0$' or '$S_K$', the eye is open during the last 1 UI period of the frame. The PSD of the FPWM is lower than NRZ at DC and higher at mid frequencies.

This spectrum imbalance of FPWM is due to different probabilities between symbols. Table II shows the probabilistic ratio between the symbol 'S$_0$' and other symbols for various FPWMs. As the pulse width resolution increases, the total number of available symbols also increases, relatively lowering the probability of 'S$_0$'. Since all symbols except 'S$_0$' have edges, an increase in the probability of these symbols boosts the mid-frequency components of the FPWM signal.

derived from the unified FPWM theory enables designing and implementing other complex FPWMs having different frame lengths and pulse width resolutions. The FPWM encoder and decoder with longer frame length than previous FPWM were simulated as an example in MATLAB. In the simulation, 14bit was coded in 8UI to have bitrate close to PAM4. The received FPWM signals were successfully decoded without bit error, and their logical feasibility was verified.

TABLE II
THE PROBABILITY OF SYMBOLS FOR VARIOUS FPWMs

| K | m | # of symbols in all possible arrays ($m \times N$) | # of S$_0$ (no edge) | # of S$_{1\sim K}$ |
|---|---|---|---|---|
| 1* | 8 | 2048 | 1024 (50%) | 1024 (50%) |
| 2 | 8 | 12776 | 5911 (46.3%) | 6865 (53.7%) |
| 3 | 8 | 47168 | 20636 (43.8%) | 26532 (56.2%) |
| 4 | 8 | 131944 | 55296 (41.9%) | 76648 (58.1%) |

* FPWM with K=1 is equivalent to NRZ.

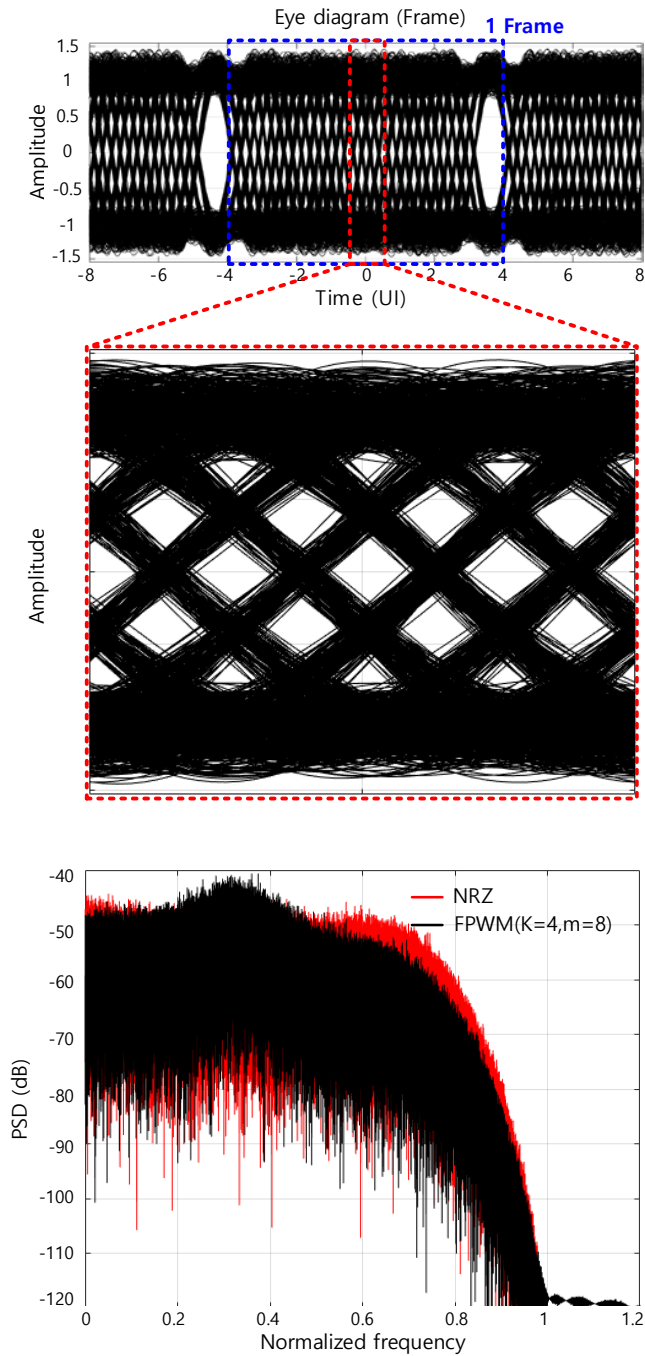

Fig. 10. Simulated eye-diagram and power spectral density

## IV. CONCLUSION

This paper proposes a general FPWM scheme and its coding algorithm. The proposed FPWM scheme shows 25% higher bitrate than the previous FPWM at the same frame length and the pulse width resolution. Meanwhile, its encoding and decoding algorithms require 93% less LUT size than the prior art. The simple coding algorithm